\documentclass[notitlepage,onecolumn,showpacs]{revtex4-1}
\usepackage[utf8]{inputenc}
\usepackage{amsmath}
\usepackage{amsfonts}
\usepackage{amssymb}
\usepackage{times,fullpage}
\usepackage{comment}
\usepackage{array,graphicx}

\graphicspath{{./figures/}}

\begin{document}

\newcommand{\nwc}{\newcommand}
\nwc{\vs}{\vspace}
\nwc{\hs}{\hspace}
\nwc{\la}{\langle}
\nwc{\ra}{\rangle}
\nwc{\nn}{\nonumber}
\nwc{\Ra}{\Rightarrow}
\nwc{\wt}{\widetilde}
\nwc{\lw}{\linewidth}
\nwc{\ft}{\frametitle}
\nwc{\dg}{\dagger}
\nwc{\mA}{\mathcal A}
\nwc{\mD}{\mathcal D}
\nwc{\mB}{\mathcal B}

\nwc{\Tr}[1]{\underset{#1}{\mbox{Tr}}~}
\nwc{\pd}[2]{\frac{\partial #1}{\partial #2}}
\nwc{\ppd}[2]{\frac{\partial^2 #1}{\partial #2^2}}
\nwc{\fd}[2]{\frac{\delta #1}{\delta #2}}
\nwc{\pr}[2]{K(i_{#1},\alpha_{#1}|i_{#2},\alpha_{#2})}
\nwc{\av}[1]{\left< #1\right>}

\nwc{\zprl}[3]{Phys. Rev. Lett. ~{\bf #1},~#2~(#3)}
\nwc{\zpre}[3]{Phys. Rev. E ~{\bf #1},~#2~(#3)}
\nwc{\zpra}[3]{Phys. Rev. A ~{\bf #1},~#2~(#3)}
\nwc{\zjsm}[3]{J. Stat. Mech. ~{\bf #1},~#2~(#3)}
\nwc{\zepjb}[3]{Eur. Phys. J. B ~{\bf #1},~#2~(#3)}
\nwc{\zrmp}[3]{Rev. Mod. Phys. ~{\bf #1},~#2~(#3)}
\nwc{\zepl}[3]{Europhys. Lett. ~{\bf #1},~#2~(#3)}
\nwc{\zjsp}[3]{J. Stat. Phys. ~{\bf #1},~#2~(#3)}
\nwc{\zptps}[3]{Prog. Theor. Phys. Suppl. ~{\bf #1},~#2~(#3)}
\nwc{\zpt}[3]{Physics Today ~{\bf #1},~#2~(#3)}
\nwc{\zap}[3]{Adv. Phys. ~{\bf #1},~#2~(#3)}
\nwc{\zjpcm}[3]{J. Phys. Condens. Matter ~{\bf #1},~#2~(#3)}
\nwc{\zjpa}[3]{J. Phys. A ~{\bf #1},~#2~(#3)}
\nwc{\zpjp}[3]{Pramana J. Phys. ~{\bf #1},~#2~(#3)}

\title{Using a Szilard engine to illustrate the validity of the modified Jarzynski equality in presence of measurement errors}
 \author{Sourabh Lahiri$^1$} 
 \email{sourabh.lahiri@icts.res.in}
  \author{A. M. Jayannavar$^{2,3}$}
 \email{jayan@iopb.res.in}
 \affiliation{$^1$ International Centre for Theoretical Sciences (TIFR), Survey no. 151, Sivakote Village, Hesaraghatta Hobli, Bengaluru 560089, India\\
   $^{2}$Institute of Physics, Sachivalaya Marg, Bhubaneswar 751005, India\\
 $^3$ Homi Bhabha National Institute, Training School Complex, Anushakti Nagar, Mumbai 400085, India}

\begin{abstract}
It has recently been shown that the Jarzynski equality gets modified, when there are experimental errors in computing work. This modified result also holds good in presence of feedback. In this work, we use a simple toy model, that of a Szilard engine, to prove these results both in the presence and in absence of feedback. 
\end{abstract}
\pacs{}

\maketitle

\section{Introduction}

The last couple of decades have observed intensive research on the so-called fluctuation theorems (FTs), which consist of a group of equalities that remain valid even when the system of interest is driven far away from their equilibrium states. Prominent among them are the Jarzynski Equality (JE) \cite{jar97_prl,jar97_pre} and the Crooks Fluctuation Theorem (CFT) \cite{cro98_jsp,cro99_pre} for work, and the detailed and integral FTs for total entropy change \cite{sei05_prl,sei08_epjb}. A detailed report on these theorems and their experimental verifications have been provided in \cite{sei12_rpp}. Other FTs have been derived for exchanged heat between two bodies \cite{jar04_prl,gom06_pre,lah14_epjb,lah16_arxiv} and for observables that do not follow exact FTs \cite{lah15_injp}. The Jarzynski equality, which will be the main focus of this article, states the following. Let us consider a system that is in contact with a thermal bath at temperature $T$.
The system is initially at equilibrium with this bath, and at time $t=0$ an external protocol $\lambda(t)$ (a parameter that is a given function of time) is switched on, and the system evolves under this protocol up to time $t=\tau$, when the protocol is switched off. One can compute the work done in this process by using the definition of work that follows from stochastic thermodynamics \cite{sek98}:
\begin{align}
  W = \int_0^\tau dt \dot\lambda \pd{H_\lambda(z)}{\lambda},
  \label{W}
\end{align}
$H_\lambda$ being the Hamiltonian of the system, and $z$ denotes the state of the system in its phase space. By performing this experiment a large number of time, an ensemble of realizations is generated. The JE states that for this ensemble, the following relation holds:
\begin{align}
  \av{e^{-\beta W}}_\Lambda=e^{-\beta\Delta F},
  \label{JE}
\end{align}
where $\beta=1/(k_BT)$, $k_B$ being the Boltzmann constant, and $\Delta F$ is the change in equilibrium free energy of the system during the process. $\Lambda\equiv \{\lambda(t)\}_0^\tau$ describes the full functional form of the protocol, and $\av{\cdots}_\Lambda$ represents ensemble averaging.

To proceed further, we need to define what is called the reverse process. In this process, the forward protocol is time-reversed, i.e. we apply the protocol $\tilde\Lambda\equiv \{\lambda(\tau-t)\}_0^\tau$. We assume that corresponding to the phase space trajectory $Z=\{z\}_0^\tau$ in the forward process described by $\Lambda$, there is a finite probability of observing the time-reversed trajectory $\tilde Z=\{\tilde z(\tau-t)\}_0^\tau$ in the reverse process described by $\tilde\Lambda$. The notation $\tilde z$ implies that the variables like velocity, which have odd parity under time-reversal, switch signs. The variables in the reverse process will henceforth be denoted by an overhead tilde symbol.  Thus, $\tilde z$ is the time-reversed state of $z$ and is a point on $\tilde Z$, $\tilde W$ is the work done along  $\tilde Z$, etc.

Now let us describe the motivation for studying relation \eqref{MJE} below. The experimental verification of the work fluctuation theorems (JE and CFT) rely on the fact that the state of the system is measured accurately and hence the work done on the system is precisely known. If there is an appreciable amount of error in the measurement, then the measured work $W_m$ can be quite different from the true work $W$ done on the system, and as a result we would observe violations of the work fluctuation theorems. In particular, we will find that
\begin{align}
  \av{e^{-\beta W_m}}\neq e^{-\beta\Delta F}.
\end{align}
In such a case, as shown in \cite{lah16_pre,wac16_arxiv}, we get the modified JE:
\begin{align}
  \av{e^{-\beta(W_m-\Delta F)}}_\Lambda = \av{e^{\beta(\tilde W_m-\tilde W)}}_{\tilde\Lambda}.
  \label{MJE}
\end{align}
In the right hand side, the averaging has been performed over trajectories $\tilde Z$ generated in the reverse process. The measured work is defined in the same way as the true work \cite{lah16_pre}:
\begin{align}
  W_m = \int_0^\tau dt\dot\lambda\pd{H_\lambda(z_m)}{\lambda}.
  \label{W_m}
\end{align}
It is obvious from the definitions of $W$ and $W_m$ that they reverse signs when calculated along the reverse phase space trajectories: $\tilde W=-W$ and $\tilde W_m=-W_m$.

\begin{figure}[!h]
  \centering
  \begin{minipage}{8cm}
    \includegraphics[width=8cm]{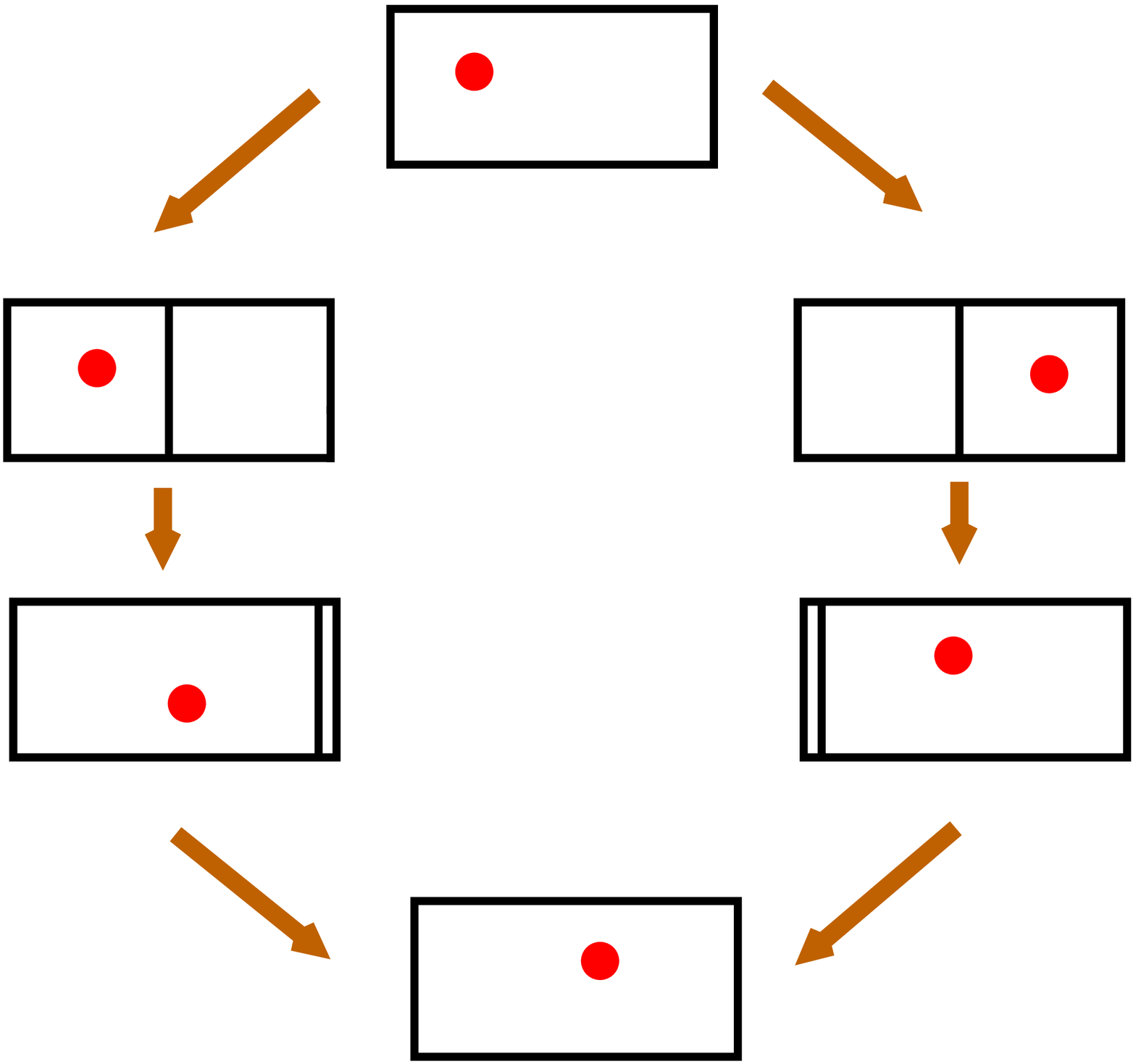}
  \end{minipage}
  \hspace{1cm}
  \begin{minipage}{2cm}
    \includegraphics[width=2cm]{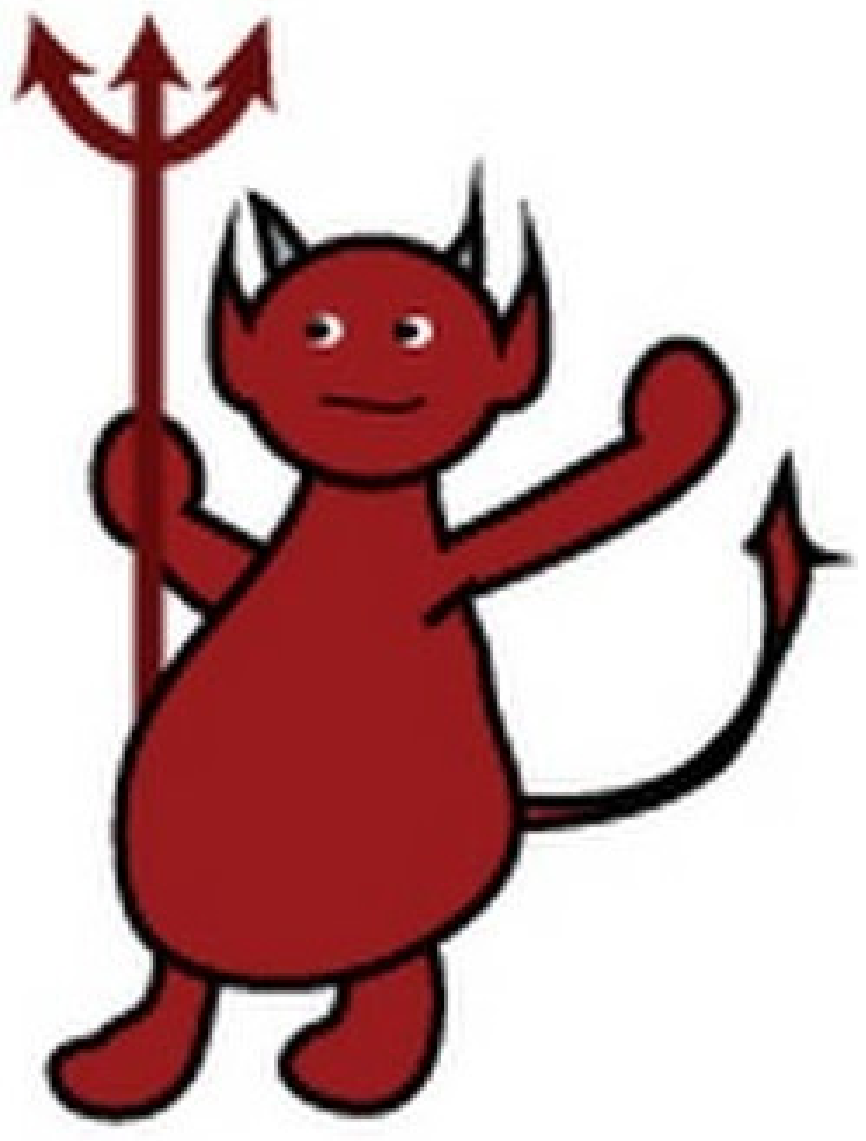}
  \end{minipage}
  \label{fig:szilard}
  \caption{Original gedanken experiment of Szilard carried out by the intelligent being or Maxwell's demon \cite{maxdem}.}
\end{figure}
Our objective in this article is to demonstrate the validity of eq. \eqref{MJE} using the very simple example of a Szilard engine.  Let us first describe the original gedanken experiment devised by Szilard (see fig. \ref{fig:szilard}). We initially have a particle that is undergoing thermal motion within a box of volume $V$ that is in thermal contact with a heat reservoir. An ``intelligent being'' now inserts a partition in the middle of the box and measures which side of the partition the particle is in. If he finds the particle to be on the left, denoted by the state $z=L$, he allows the partition to quasistatically move towards right till it touches the right wall of the box. At the end of this process, he removes the partition and the particle now comes back to its initial state. Similarly, if he finds the particle to be on the right of the partition, denoted by $z=R$, he allows the partition to quasistatically move towards left till it touches the left wall of the box, and finally removes the partition. Thus, in this process, the initial volume within which the particle is confined is $V_i=V/2$ and the final volume is $V_f=V$. The external parameter in the experiment is the volume itself, and the work done on the particle is given by
\begin{align}
  W &= \int_{0}^{\tau} dt\dot V\pd{H}{V}=-\int_{V_i}^{V_f}pdV = -k_BT\ln\frac{V_f}{V_i}\nn\\
  &= -k_BT\ln\frac{V}{V/2} = -k_BT\ln 2.
\end{align}
Here, the process being quasistatic, the time of observation $\tau\to\infty$
As a result, an amount of work equal to $k_BT\ln 2$ is always extracted in the process, simply by using the thermal motion of the particle. This apparently looks to be in contradiction with the second law of thermodynamics. However, Landauer had argued that the full cycle can only be completed on erasure of the memory of this intelligent being, and that this erasure process entails at least $k_BT\ln 2$ amount of work that needs to be done by him. This is how the Maxwell's demon is exorcised by the Landauer's erasure principle. The problem has been attacked with a different point of view by Sagawa and Ueda \cite{sag10_prl} and later on by others \cite{pon10_pre,lah12_jpa} (see \cite{sag12_pre} and the references therein). It has been shown that if we exclude the memory device (or the memory of the intelligent being) from our analysis, the system of interest must follow a modified second law that involves the so-called average mutual information between the measured and the actual states. Introduction of this correction term once again saves the second law.

Let us now use this simple setup to illustrate eq. \eqref{MJE}.

\section{Verification of modified work relation in absence of feedback}

\begin{figure}[!h]
  \centering
  \includegraphics[width=8cm]{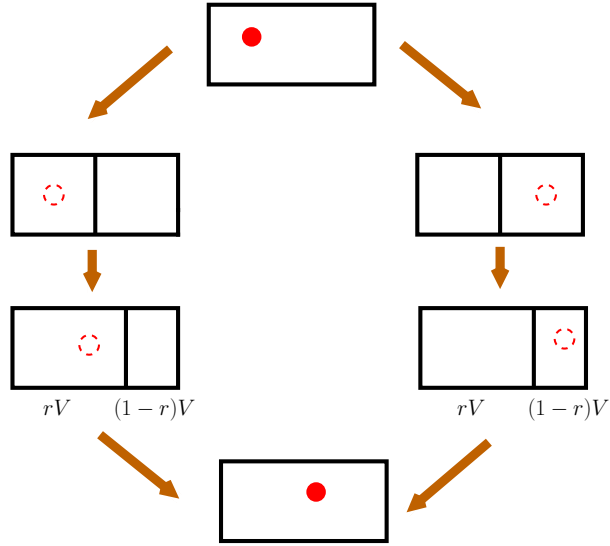}
  \caption{Our experiment on the Szilard engine in absence of any feedback, so that the partition is moved towards right, irrespective of the outcome of the measurement. The dotted circles represent the measured state of the particle, which can be different from the true state.}
  \label{fig:setup}
\end{figure}
We depict our experiment in fig. \ref{fig:setup}.
We have a particle inside a box that is in thermal contact with its environment. The particle is at thermal equilibrium with its environment. We insert a partition in the middle of the box and measure the state of the particle, i.e. whether it is on the left ($z_m=L$) or on the right ($z_m=R$) of the partition, where $z_m$ denotes the measured state, that can be different from the true state $z$ of the particle. Irrespective of what the result of the measurement is, we quasistatically move the partition towards right till it divides the total volume $V$ in the ratio $r:(1-r)$, where $1/2<r<1$. In other words, at this stage the volume on the left of the partition is $rV$ and on the right of the partition is $(1-r)V$. At the end of the process, the partition is removed, so that the state of the system is the same as that at the beginning of the process. The true work is given by
\begin{align}
 W =  -k_BT\ln\frac{V_f}{V_i},
\end{align}
where $V_i$ and $V_f$ are the initial and final volumes in which the particle is {\it actually} confined. The measured work, on the other hand, is given by
\begin{align}
  W_m = -k_BT\ln\frac{V_{fm}}{V_{im}},
\end{align}
where $V_{im}$ and $V_{fm}$ are the initial and final volumes in which the particle is {\it known} to be confined through the measurement. The measured position $z_m$ can be equal to the true position $z$ with probability $q$ or different with probability $(1-q)$. We need to verify the equality
\begin{align}
  \av{e^{-\beta(W_m-\Delta F)}}_{\Lambda} = \av{e^{\beta(\tilde W_m-\tilde W)}}_{\tilde \Lambda},
  \label{MFT}
\end{align}
where $W_m$ is the measured work, $\tilde W_m$ and $\tilde W$ are the measured and the true works in the reverse process, respectively. 
In the reverse process, the partition is inserted so as to divide the total volume of the box in the ratio $r:(1-r)$, and then it is moved quasistatically to the middle of the box.

Note that if the particle is measured to be on the left in the forward process ($z_m=L$), then $V_{im}=V/2$ and $V_{fm}=rV$, while if $z_m=R$ then $V_{im}=V/2$ but $V_{fm}=(1-r)V$.
Since $\Delta F=0$ in the process (initial and final states of the particle are same), the LHS of the above equation is given by
\begin{align}
  \av{e^{-\beta W_m}} &= \frac{1}{2}e^{-\beta(-k_BT\ln[2r])}+\frac{1}{2}e^{-\beta(-k_BT\ln[2(1-r)])}\nn\\
                      &= r+1-r =1. 
\end{align}
Here, we have used the fact that the probability of observing the particle to be on the left or on the right are equal to 1/2.

To calculate the RHS, the various cases that can be considered are
\begin{enumerate}
\item $z_m=z=L$ : ~~~In this case, both $W$ and $W_m$ equal $-k_BT\ln[2r]$, so that $\tilde W_{(1)} = k_BT\ln[2r] = \tilde W_{m(1)}$.  The measured outcome is ``correct'' ($\tilde z_m = \tilde z$) with probability $q$. Further, the probability of  $\tilde z$ being equal to $L$ is $r$, since initially the partition divides the volume in the ratio $r:(1-r)$. Thus, the net probability of observing the above mentioned values of $\tilde W$ and $\tilde W_m$ is equal to $qr$.

\item $z_m=L, ~z=R$ : ~~~In this case, we have $\tilde W_{(2)} = k_BT\ln[2(1-r)]$, ~~$\tilde W_{m(2)} = k_BT\ln[2r]$, which are obtained by reversing the signs of $W$ and $W_m$ obtained in the forward process. Since probability of a wrong measurement ($\tilde z_m\neq\tilde z$) is $(1-q)$ and that of having $\tilde z$ equal to $R$ is $(1-r)$, the net probability of observing the above mentioned values of $\tilde W$ and $\tilde W_m$ is $(1-q)(1-r)$.

\item $z_m=R, z=L$ : ~~~$W_{(3)}^\dg = k_BT\ln[2r]$, ~~$W_{m(3)}^\dg = k_BT\ln[2(1-r)]$, with probability $(1-q)r$.

\item $z_m=z=R$ : ~~~$W^\dg_{(4)} = k_BT\ln[2(1-r)] = W^\dg_{m(4)}$ with probability $q(1-r)$.
\end{enumerate}
Thus, the RHS of eq. \eqref{MFT} gives
\begin{align}
  \av{e^{\beta(\tilde W_m-\tilde W)}} &= qr + (1-q)(1-r)\exp\bigg[\ln\frac{r}{1-r}\bigg]\nn\\
                                &~~~~~~+(1-q)r\exp\bigg[\ln\frac{1-r}{r}\bigg]+q(1-r)\nn\\
                                &=q+(1-q)r+(1-q)(1-r) = 1.
\end{align}
Thus we have verified that the LHS of \eqref{MFT} equals its RHS and is equal to 1.

\section{Verification of modified work relation in presence of feedback}

\begin{figure}[!h]
  \centering
  \includegraphics[width=8cm]{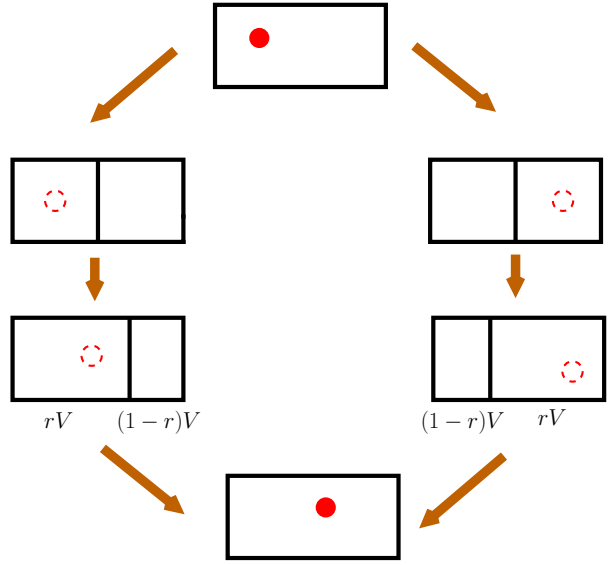}
  \caption{Our experiment on the Szilard engine in presence of feedback, where the partition is moved so as increase the volume in which the particle is known (with some error) to be confined. The dotted circles represent the measured state of the particle, which can be different from the true state.}
  \label{fig:setup_feedback}
\end{figure}

We consider the following process (see fig. \ref{fig:setup_feedback}): as before, the particle is initially present in a box that is in thermal contact with its environment. We insert a partition at the middle of the box and measure the state of the particle (whether it is on the left or on the right of the partition). If $z_m=L$, then we move the partition quasistatically towards right till the total volume gets divided in the ratio $r:(1-r)$.  If $z_m=R$, then the partition is moved quasistatically towards left till the ratio is $(1-r):r$. 
 Since $V_{im}=V/2$ and $V_{fm}=rV$ for the forward process, we have
\begin{align}
  W_m = -k_BT\ln\frac{V_{fm}}{V_{im}} = -k_BT\ln[2r],
  \label{Wm_feedback}
\end{align}
irrespective of the value of $z_m$. As shown in \cite{wac16_arxiv}, the same equality still holds:
\begin{align}
  \av{e^{-\beta(W_m-\Delta F)}}_\Lambda = \av{e^{\beta(\tilde W_m-\tilde W)}}_{\tilde\Lambda}.
  \label{MJE_feedback}
\end{align}
It is important to note that in presence of feedback, $\tilde\Lambda$ is defined such that it depends on the measured outcomes in the forward process \cite{hor10_pre,lah12_jpa}.
The LHS of \eqref{MJE_feedback} becomes (using $\Delta F=0$)
\begin{align}
  \av{e^{-\beta(-k_BT\ln[2r])}} = 2r.
\end{align}

Let the probability of correct measurement be $q$ and that of a wrong measurement be $1-q$.
The following cases can occur:
\begin{enumerate}
\item $z_m=z=L$ : ~~~$\tilde W_{(1)} = k_BT\ln[2r] = \tilde W_{m(1)}$ with probability $qr$ (since probability of $\tilde z_m= \tilde z$ is $q$, and of $\tilde z=L$ is $r$).

\item $z_m=L, ~z=R$ : ~~~$W_{(2)}^\dg = k_BT\ln[2(1-r)]$, ~~$W_{m(2)}^\dg = k_BT\ln[2r]$, with probability $(1-q)(1-r)$. This is because the probability of $\tilde z_m\neq \tilde z$ is $(1-q)$ and that of $\tilde z=R$ is $(1-r)$.

\item $z_m=R, ~z=L$ : ~~~$W_{(3)}^\dg = k_BT\ln[2(1-r)]$, ~~$W_{m(3)}^\dg = k_BT\ln[2r]$ with probability $(1-q)(1-r)$. Note the difference of this case and the next one from the same cases in absence of feedback.

\item $z_m=z=R$ : ~~~ $W_{(4)}^\dg = k_BT\ln[2r] = W_{m(4)}^\dg$ with probability $qr$.
\end{enumerate}

Therefore, the RHS of \eqref{MJE_feedback} gives
\begin{align}
  \av{e^{\beta(\tilde W_m-\tilde W)}} &= 2\times\bigg[qr\exp(0)+(1-q)(1-r)\exp\bigg(\ln\frac{r}{1-r}\bigg)\bigg] \nn\\
                                      &= 2[qr+(1-q)r] = 2r.
                                        \label{RHS_feedback}
\end{align}
Thus, eq. \eqref{MJE_feedback} is verified. Let us check what happens in Szilard's original experiment. In this case, $q=1$ and $r=1$, so that from \eqref{MJE_feedback} and \eqref{RHS_feedback}, we obtain the equality
\begin{align}
  \av{e^{-\beta(W-\Delta F)}}=2,
\end{align}
which shows that the efficacy parameter for Szilard's engine is equal to 2 \cite{sag10_prl}.

\section{Conclusions}

In this work, we have used a simple pedagogical setup, that of a Szilard engine, to demosntrate the validity of the modified Jarzynski equality (see Eq. \eqref{MJE} above) in presence of measurement errors. We begin with an experiment where there are faulty measurements of the system's state but no feedback is applied \cite{lah16_pre,wac16_arxiv}. Later, in accordance with the findings of \cite{wac16_arxiv}, we show that the relation is true even when feedback is applied, based on the faulty measurements. We hope that this simple example would serve to provide a nice visualization and understanding of this relation.

\section{Acknowledgement}

One of us (AMJ) thanks DST, India for awarding J. C. Bose National Fellowship. SL thanks Deepak Bhat for useful discussions.


%

\end{document}